\documentclass[aps,pra,twocolumn]{revtex4}
\usepackage{graphicx}
\begin{document}
\title{Thermal entanglement in the nanotubular system $\rm {\bf Na_2V_3O_7}$ }
\author{T. V\'ertesi$^1$ and E. Bene$^2$}
\affiliation{$^1$ Institute of Nuclear Research of the Hungarian Academy of
Sciences,\\ H-4001 Debrecen, P.O. Box 51, Hungary} \affiliation{$^2$
Institute of Chemistry, Chemical Research Center, Hungarian Academy of
Sciences, Budapest, Hungary}

\date{\today}

\begin{abstract}
Macroscopic entanglement witnesses have been put forward recently to reveal
nonlocal quantum correlations between individual constituents of the solid
at nonzero temperatures. Here we apply a recently proposed universal
entanglement witness, the magnetic susceptibility [New J. Phys. {\bf 7}, 258
(2005)] for the estimation of the critical temperature $T_c$ in the
nanotubular system ${\rm Na_2V_3O_7}$ below which thermal entanglement is
present. As a result of an analysis based on the experimental data for
dc-magnetic susceptibility, we show that $T_c \approx 365$~K, which is
approximately three times higher than the critical temperature corresponding
to the bipartite entanglement.
\end{abstract}
\maketitle

\noindent\textbf{I. Introduction}
\\Entanglement lies in the heart
of quantum mechanics, revealing the existence of nonlocal correlations that
cannot be described by classical physics alone \cite{Peres93}. Because of
such strong correlations over and above the classical one, entangled states
have been recognized as a crucial resource for quantum information
processing \cite{Nielsen00}. Thus for, experimental demonstrations of
entanglement have been confined yet mostly to atomic scales. This is due to
the subtle nature of quantum entanglement, since extending the size of the
physical object increases the number of constituents, and the number of
degrees of freedom that can interact with the environment, which is
responsible for reducing and eventually destroying entanglement in the
system.

However, it was demonstrated recently that entanglement can affect the
macroscopic properties (such as heat capacity or magnetic susceptibility) of
an insulating magnetic compound ${\rm LiHo_xY_{1-x}F_4}$ \cite{Ghosh03}, but
at a very low (below 1~K) temperature. The basic reason for it is that
magnetic susceptibility is proportional to the two-site correlations between
the spins in the magnetic solid, which can be higher than the one allowed by
classical physics alone, when entanglement is present in the material as
well. In the light of the result of Ghosh et al \cite{Ghosh03} that
entanglement can have a significant effect in the macroscopic world, the
question arises whether we can experience entanglement at higher
temperatures as well.

As a next step, Brukner et al \cite{BVZ04} showed that for uncoupled dimers
at zero magnetic field the magnetic susceptibility serves as an entanglement
witness (EW), i.e., an observable which can distinguish between entangled
and separable states in a quantum system \cite{HHH96}. It has been
demonstrated \cite{BVZ04} that the experimental data on magnetic
susceptibility in the spin-1/2 alternating bond antiferromagnet CN ${\rm
(Cu{(NO_3)}_22.5D_2O)}$ \cite{BFS63} implies the presence of entanglement
below the temperature 5 Kelvin.

On the other hand, Bose and Tribedi \cite{BT05} showed, based on energy
considerations \cite{Toth05}, that the magnetic susceptibility as an EW can
be applied to compounds which can be modeled as a collection of independent
clusters of $N$ spins, when the Hamiltonian is isotropic in spin space.  As
a special case, Bose and Tribedi considered the $S=1/2$ polyoxovanadate AFM
compound V12, which material can be treated as a collection of independent
$S=1/2$ tetramers and estimated that the critical temperature below which
entanglement exists in the system is $T_c \approx 25.4$~K.

Other thermodynamical observables, such as internal energy, magnetization,
or heat capacity \cite{Wang05,DDB04,BV04,WBSL04,Toth05,WVB05b}, have also
been proposed for EW but either they rely on a model Hamiltonian of the
system (and usually not a directly measurable quantity) or are applicable
only for relatively low temperatures \cite{WVB05b}.

However, recently Wie\'sniak et al \cite{WVB05a} constructed an EW based on
the sum of magnetic susceptibilities measured along the three orthogonal
spatial directions. This entanglement witness is applicable to systems
without full knowledge of the specific model of the solid. The separability
criterion of Wie\'sniak et al \cite{WVB05a} for $N$ spin-$l$ particles reads
as follows:
\begin{eqnarray}
\label{in2}
\bar \chi  = \chi _x  + \chi _y  + \chi _z  \ge \frac{C}{T}l\;,
\end{eqnarray}
with $ C = \left( {g\mu _B } \right)^2 N/k$, where $g$ is the $g$-factor,
$\mu_B$ is the Bohr magneton, $k$ is the Boltzmann constant, and $N$ denotes
the number of particles in the volume. Further in the right-hand side of
(\ref{in2}), $\bar \chi$ denotes the sum of magnetic susceptibilities per
volume measured along the orthogonal $x,y$, and $z$ directions. In the
isotropic case for spin-$1/2$ particles inequality~(\ref{in2}) simplifies to
the inequality
\begin{equation}
\label{in1} \chi _z  \ge \frac{C}{T}\frac{1}{6}{\rm}\;.
\end{equation}
Therefore, if $\chi_z < \frac{C}{T}\frac{1}{6}{\rm}$ the isotropic solid
state system contains entanglement between the individual spin-$1/2$
systems. Note, that $\chi_z$ entanglement witness in (\ref{in1}) is the same
as the one in Ref.~\cite{BVZ04}, although that has been introduced only for
uncoupled dimers using a simpler argumentation. Since the entanglement
witness $\bar \chi$ of Wie\'sniak et al \cite{WVB05a} is completely general,
the entanglement criterion~(\ref{in1}) can be applied for a large class of
quantum spin systems, as it neither rely on the existence of small spin
clusters or on the full knowledge of the system's Hamiltonian.

The low-dimensional quantum magnets have recently been in the focus of both
experimental and theoretical research. A large class of these materials are
the spin-$1/2$ quantum magnets with chain, ladder, or planar geometries. In
this respect new possibilities emerged, when Millet et al \cite{Millet99}
recognized that the variation of some two-dimensional compounds results in a
nanotubular structure, and as a result the first
transition-metal-oxide-based nanotubular system ${\rm Na_2V_3O_7}$ was
synthesized. In this system the ${\rm VO_5}$ pyramids share edges and
corners to form a nanotubular structure with ${\rm Na}$ atoms located inside
each individual tube. The ab initio analysis in Ref.~\cite{Dasgupta04}
showed that ${\rm Na_2V_3O_7}$ can be described as formed by nanotubes
consisting of weakly coupled nine-site rings.

The main purpose of the present paper is to demonstrate that the
entanglement witness $\chi_z$ in criterion~(\ref{in1}) implies entangled
states in the system ${\rm Na_2V_3O_7}$ at the room temperature range, and
we estimate the critical temperature $T_c^{exp} \approx 365$~K. The validity
of this result is also supported by additional calculations. In particular
we find that neglecting the inter-ring couplings as a first approximation,
the inequality $ \left| {\left\langle {{\bf S}_i {\bf S}_j } \right\rangle }
\right| \le 1/4 $, where $i,j$ denotes spin sites within the ring, may
indicate pairwise entanglement. Based on the parameters of the model
Hamiltonian of the system ${\rm Na_2V_3O_7}$ which are derived from ab
initio molecular structure calculations \cite{Dasgupta04}, we calculate the
critical temperature corresponding to $ \left| {\left\langle {{\bf S}_i {\bf
S}_j } \right\rangle } \right| = 1/4 $, and obtain $ T_c^{pair} \approx 123
$~K, below which pairwise entanglement is present in the system. The three
times smaller temperature bound for the bipartite entanglement indicates
that a large portion of entanglement is stored in correlations between more
than two spins within the nine-site ring. Therefore we conclude that the
entanglement above the temperature $ T_c^{pair} $ is multipartite in nature.
To account for this effect we analyze the symmetric trimer (described by
$S=1/2$ Heisenberg antiferromagnetic (AFM) isotropic Hamiltonian), since by
small temperatures the nine-spin cluster model of the ${\rm Na_2V_3O_7}$
system corresponds to a three-site ring. Although the symmetric trimer is
known not to have pairwise entanglement both at $T=0$ and at finite
temperatures \cite{Wang02}, we find multipartite entanglement witnessed by
criterion~(\ref{in1}) in the system. We argue that the "missing
entanglement" in the case of the symmetric trimer, not explored by the $
\left| {\left\langle {{\bf S}_i {\bf S}_j } \right\rangle } \right| \le 1/4$
condition, appears in the nanotubular system ${\rm Na_2V_3O_7}$ as well, and
the existence of this multipartite entanglement causes the large difference
in the values of critical temperatures $T_c^{exp}$ and $T_c^{pair}$.
Further, we claim that due to the weakly coupled nine-site rings with
respect to the intraring couplings, the entangled state itself in the
nanotubular system is shared with a good approximation among the nine spins
within a ring. Therefore the entangled state is confined to the rings of the
nanotube, and does not extend to macroscopic sizes. This fact may partially
explain why the entanglement in this system persists up to such high
temperatures.

\smallskip

\noindent \textbf{II. Entanglement witnesses based on two-site
spin correlation functions}\\
When a system is in its thermal equilibrium under a certain temperature $T$,
it can be described by the density operator $ \rho \left( T \right) = Z^{ -
1} \exp \left( { - \beta H} \right) $, where $H$ is the Hamiltonian, $Z$ is
the partition function, and $ \beta  = 1/kT$. A thermal state remains
entangled up to a critical temperature $T_c$ above which the state becomes
separable, i.e., the amount of entanglement vanishes.

In order to measure the thermal entanglement based on the thermal density
matrix we need to know both the eigenvalues and eigenvectors of the
Hamiltonian. On the other hand, bulk properties of solids such as internal
energy $ \left( {U =  - Z^{ - 1} \frac{{\partial Z}}{{\partial B_z}}}
\right)$, magnetization $ \left( {\left\langle M_z \right\rangle = -
(Z\beta)^{ - 1} \frac{{\partial Z}}{{\partial {B_z} }}} \right)$, or
magnetic susceptibility $\left( {\chi_z  = \frac{{\partial \left\langle M_z
\right\rangle }}{{\partial B_z}}} \right)$, can be derived from the
partition function alone; thus these quantities are defined merely by the
eigenvalues of the system $H$.

The latter relationship between magnetization and magnetic susceptibility
has given rise to the notion of thermodynamical complementarity \cite{HV05}.
On the other hand, if $[M,H]=0$ the equality
\begin{eqnarray}
\label{in3} \chi _z  = \frac{C}{NT}\Delta ^2 M_z  =
\frac{C}{NT}\left( {\left\langle {M_z^2 } \right\rangle  -
\left\langle {M_z } \right\rangle ^2 } \right)
\end{eqnarray}
holds linking the fluctuation of magnetization to the macroscopic observable
$\chi_z$. Here $\Delta ^2 M_z$ denotes the variance of the magnetization,
while the constant $C$ has the same meaning as in inequality~(\ref{in2}) and
$N$ is the number of particles in the volume. On the other hand, noticing
that $ M_z  =\sum\limits_{i = 1}^N {S_i^z} $, formula~(\ref{in3}) also
connects the magnetic susceptibility to its microscopic origin in the form
of spin-spin correlation functions.

Now we represent two entanglement witnesses from the literature, based on
spin-spin correlation functions, which are crucial to construct macroscopic
entanglement witness from the magnetic susceptibility through
equation~(\ref{in3}). This promise to reveal thermal entanglement in
macroscopic systems even when the eigenvectors of the system's Hamiltonian
are not known. For any separable states of $N$ spins of length $l$, that is,
for any convex sum of $N$ product states, $ \rho  = \sum\limits_k {\omega _k
\rho _k^1 \otimes } \rho _k^2  \otimes \ldots \otimes \rho _k^N$, one has
\cite{BV04}
\begin{eqnarray}
\label{in8}
 \sum\limits_{i = 1}^N {\left| {\left\langle {{\bf S}_i
{\bf S}_{i + 1} } \right\rangle } \right|}  \le Nl^2\;,
\end{eqnarray}
which especially for two qubits (at sites $i$ and $j$) yields the
inequality
\begin{eqnarray}
\label{in9}
 \left| {\left\langle {{\bf S}_i {\bf S}_j }
\right\rangle } \right| \le \frac{1}{4}\;.
\end{eqnarray}
The other inequality for separable states of $N$ spins of length $l$ is
based on entanglement detection applying the sum uncertainty relations
\cite{Hofmann03,WVB05a},
\begin{eqnarray}
\label{in10}
 \Delta ^2 M_x  + \Delta ^2 M_y  + \Delta ^2 M_z  \ge Nl\;.
\end{eqnarray}
Now we restrict ourselves to Heisenberg spin-1/2 lattices at zero field with
isotropic, but in general inhomogeneuous coupling constants $J_{ij}$,
\begin{eqnarray}
\label{in11}
 H = \sum\limits_{i \ne j} {J_{ij} {{\bf
S}_i {\bf S}_j } }\;.
\end{eqnarray}
Due to the isotropy of the Hamiltonian in spin space, the magnetization
vanishes in all three orthogonal directions, and thus $\Delta ^2 M_\alpha =
\left\langle {M_\alpha ^2 } \right\rangle$ for $\alpha=x,y,z$. Using the
relation $ M_\alpha  =\sum\limits_{i = 1}^N {S_i^\alpha } $ and assuming
spin-1/2 lattices Eq.~(\ref{in10}) simplifies to
\begin{eqnarray}
\label{in12}
 \sum\limits_{i,j = 1}^N {\left\langle {{\bf S}_i {\bf
S}_j } \right\rangle }  \ge \frac{N}{2}\;,
\end{eqnarray}

From now on for brevity we denote the entanglement witness corresponding to
criterion~(\ref{in9}) by EW1, and by EW2 of the criterion~(\ref{in12}). Note
that combining inequality~(\ref{in10}) with (\ref{in3}) implies the
criterion~(\ref{in2}) (see Wie\'sniak et al \cite{WVB05a}). On the other
hand for the isotropic Heisenberg Hamiltonian~(\ref{in11}) magnetization $M$
commutes with the Hamiltonian $H$, $[M,H]=0$, thus the witness $\chi_z$ in
criterion~(\ref{in1}) is valid.  As a consequence the power of EW2 in
(\ref{in12}) to signal entanglement is equivalent with the witness $\chi_z$
in criterion~(\ref{in1}).

Wang and Zanardi \cite{WZ02} showed that for the Heisenberg
Hamiltonian~(\ref{in11}) the concurrence \cite{Wootters98}, which gives a
measure of pairwise entanglement between qubits at site $i$ and site $j$,
becomes $ C_{ij}  = 1/2 \max \left[ {0,2\left| {\left\langle {S_i^z S_j^z }
\right\rangle } \right| - \left\langle {S_i^z S_j^z } \right\rangle  - 1/4}
\right]$. The concurrence $C_{ij}=0$ corresponds to an unentangled state.
Regarding to the isotropy of the model the $\alpha=x,y$ and $z$ components
of the spin-spin correlation functions are all equal, and assuming that
$\left\langle {{\bf S}_i {\bf S}_j } \right\rangle$ is negative, we arrive
at $ C_{ij} = 1/2 \max \left[ {0, - \left\langle {{\bf S}_i {\bf S}_j }
\right\rangle  - 1/4} \right]$. Thus, whenever $\left\langle {{\bf S}_i {\bf
S}_j } \right\rangle  \le 0$ holds, pairwise entanglement between qubits $i$
and $j$ vanishes for $ \left| {\left\langle {{\bf S}_i {\bf S}_j }
\right\rangle } \right| \le 1/4$. Since the condition above is just the
separability criterion~(\ref{in9}), this implies that EW1 detects pairwise
entanglement between any two qubits of a system with
Hamiltonian~(\ref{in11}) when the corresponding spin-spin correlation has
negative value.

Now we compare the power of EW1 and EW2 detecting entanglement in a
three-qubit ring interacting via the AFM isotropic Heisenberg Hamiltonian
and show that EW2 can be more effective. For the AFM symmetric trimer the
Hamiltonian is given by $ H = J\left( {{\bf S}_1 {\bf S}_2  + {\bf S}_2 {\bf
S}_3  + {\bf S}_3 {\bf S}_1 } \right)$, with $J>0$ thus
criterion~(\ref{in12}) gives $ \sum\limits_{i,j = 1}^3 {\left\langle {{\bf
S}_i {\bf S}_j } \right\rangle }  = 3\left( {\left\langle {{\bf S}_i^2 }
\right\rangle  + 2\left\langle {{\bf S}_i {\bf S}_{i + 1} } \right\rangle }
\right) \ge \frac{3}{2}$ and consequently $ \left\langle {{\bf S}_i {\bf
S}_{i + 1} } \right\rangle  \ge - 1/8$, for which the AFM interaction
implies the relation $ \left| {\left\langle {{\bf S}_i {\bf S}_{i+1} }
\right\rangle } \right| \le 1/8$ for separable states. Thus in this case the
condition $ \left| {\left\langle {{\bf S}_i {\bf S}_{i+1} } \right\rangle }
\right| \le 1/8$ of EW2 obviously implies a lower bound on the separability
than the condition $ \left| {\left\langle {{\bf S}_i {\bf S}_{i+1} }
\right\rangle } \right| \le 1/4$ of EW1. Since direct calculation shows that
for the ground state of the symmetric trimer $\left\langle {{\bf S}_i {\bf
S}_{i + 1} } \right\rangle  = - 1/4$, EW2 indicates entangled ground state,
whereas EW1 could not reveal entanglement. The result that there exists
entanglement in the symmetric trimer over the two-site entanglement is
supported by the definition of 1-tangle $\tau_1$ as well \cite{CKW00}, which
quantifies the entanglement of one site with the rest of the chain at $T=0$,
and is given as $\tau _{\rm 1} = 4\det \left( {\rho _1 } \right)$, where
$\rho _1  = \left( {I + \sum\limits_\alpha  { \left\langle
{S_1^\alpha}\right\rangle \sigma ^\alpha } } \right)/2$ is the one-site
reduced density matrix, and $\sigma ^\alpha$ are the Pauli matrices with
$\alpha  = x,y,z$. In terms of spin expectation values
$\left\langle{S_1^\alpha}\right\rangle$, $\tau_1$ takes the form $ \tau _1 =
1 - 4\sum\limits_\alpha  {\left\langle {S_1^\alpha  } \right\rangle} ^2 $.
The vanishing of $\tau_1$ implies that there is no entanglement in the
ground state, i.e. the state is factorized. The isotropy of the symmetric
trimer $ \left\langle {S_1^x } \right\rangle  = \left\langle {S_1^y }
\right\rangle  = \left\langle {S_1^z } \right\rangle = 0$ entails that
$\tau_1=1$, which means that indeed the ground state of the symmetric trimer
is entangled. Although in the above analysis we treated merely the three
qubit isotropic Heisenberg ring, we conjecture that the results obtained
above can be generalized to systems with general Heisenberg
Hamiltonian~(\ref{in11}) for cluster size $N>3$ and for different but
positive $J_{ij}$ values as well. We intend to support this conjecture in
the next sections with the analysis of the quantum magnet $\rm Na_2V_3O_7$,
which can be considered as a nine-leg nanotubular system, and is well
described by the Hamiltonian~(\ref{in11}) with $ J_{ij}> 0 $ exchange
constants.

\smallskip

\noindent\textbf{III. Magnetic susceptibility of the quantum magnet ${\rm
\bf{Na_2V_3O_7}}$} \\$\rm Na_2V_3O_7$ is a material whose structure may be
considered as composed by nanotubes oriented along the $c$ axis of the
crystal lattice. A detailed description of the structure of this compound
can be found in Ref.~\cite{Millet99}. Following the discussion and ab initio
results given in Ref.~\cite{Dasgupta04} we construct the model Hamiltonian
of the system $\rm Na_2V_3O_7$, which is given by
\begin{eqnarray}
\label{in14}
 H = \left( {\sum\limits_{i = 1}^9 {J_1
{\bf S}_i {\bf S}_{i + 1} + } J_2^i {\bf S}_i {\bf S}_{i + 2} }
\right)\;,
\end{eqnarray}
where periodic boundary conditions were imposed $({\bf S}_{i + 9} = {\bf
S}_i)$, and the weak inter-ring couplings as a first approximation were
neglected in the model. Thus in this model we treat the system as the
collection of independent nine-site rings. The optimal parameters,
neglecting also the differences in between the nearest neighbor (NN) and in
between the next nearest neighbor (NNN) intraring interactions, are given by
$J_1=200$~K and $J_2=140$~K \cite{Dasgupta04}. Due to the structure of the
material \cite{Millet99,Dasgupta04} the ring is partially frustrated, where
$J_2^i=0$ for $i=1,4,7$ and $J_2^i=J_2$ for the indices $i=2,3,5,6,8,9$. The
two kinds of couplings in the frustrated model are depicted in Fig.~1.

%===========================================================================
\begin{figure}
\vspace{0cm}
\includegraphics[width=5cm]{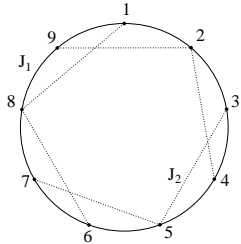}\vspace*{-3cm}
\vspace{2cm} \caption{Illustration of exchange couplings $J_1$, $J_2$ in the
spin-1/2 nine-leg ring model of the nanotube $\rm Na_2V_3O_7$. }
\label{structure}
\end{figure}
%===========================================================================

The ground state of the model is fourfold degenerate, composed of two
doublets, corresponding to the right and left chirality \cite{LNMKM04}.
Since this model is invariant under translations $i\rightarrow i+3$ it
corresponds by low energies to a three-site ring. In Fig.~2 we display the
experimentally observed inverse magnetic susceptibility of the $\rm
Na_2V_3O_7$ at temperatures between $1.9$~K and $315$~K (data is taken from
Ref.~\cite{GRMOMM}). As a comparison we also reproduced and plotted in
Fig.~2 the calculated inverse magnetic susceptibility for the partially
frustrated model from Ref.~\cite{Dasgupta04} with couplings $J_1=200$~K and
$J_2=140$~K in the temperature range $T=0-600$~K. The temperature dependence
of the magnetic susceptibility was obtained in Ref.~\cite{Dasgupta04} from
the thermal state $\rho(T)$ by exact diagonalization of the
Hamitonian~(\ref{in14}). The nice agreement between the experimental and
calculated values supports the validity of the ab initio derived parameters
$J_1$ and $J_2$ in the model.

\smallskip

\noindent\textbf{IV. Calculation of thermal entanglement in the system ${\rm
\bf{Na_2V_3O_7}}$} \\In this section we give a lower bound on the critical
temperature below which the thermal state of the system $\rm Na_2V_3O_7$ is
entangled. The two estimations for this temperature limit will be calculated
based on entanglement witnesses EW1 and EW2 (in criterions~(\ref{in9}) and
(\ref{in12}), respectively). EW1 is able to reveal pairwise entanglement
between individual spins, while EW2 may reveal residual entanglement which
is not detected by EW1, as it was demonstrated in the example of symmetric
trimer. The critical temperature related to the pairwise entanglement and
detected by EW1 will be called $T_c^{pair}$, while the possibly higher
critical temperature indicated by EW2 is denoted by $T_c^{exp}$.
%%%%%%%%%%%%%%%%%%%%%%%%%%%%%
\begin{figure}
\vspace{-1cm}
\includegraphics[width=8.7cm]{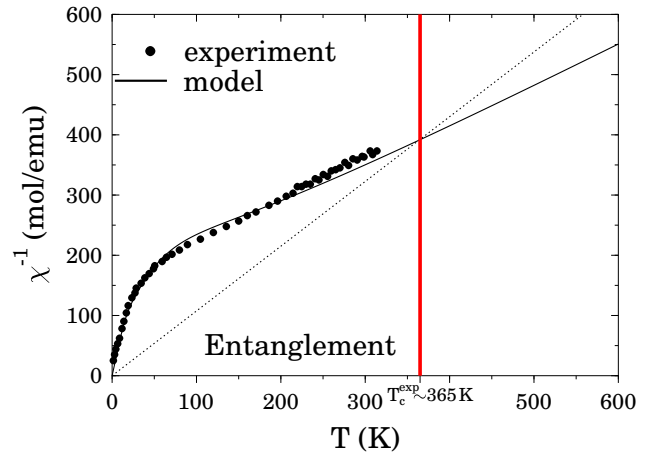}\vspace*{-2cm}
\vspace{1.5cm} \caption{Temperature dependence of the inverse magnetic
susceptibility in units of mol/emu of the system $\rm Na_2V_3O_7$. The
filled dots correspond to the experimental data from Ref.~\cite{GRMOMM}. The
solid curve is the calculated curve by reproducing the results of
Ref.~\cite{Dasgupta04}, and the dotted line represents the macroscopic
entanglement witness $\chi_z$. The vertical solid line shows the critical
temperature $ T_c^{exp} \approx 365$~K, i.e., the intersection point of the
dotted and solid curves,  below which entanglement exists in the nanotubular
system $\rm Na_2V_3O_7$ . } \label{structure}
\end{figure}
The relation~(\ref{in3}) between magnetic susceptibility and two-site spin
correlation functions links EW2 to the entanglement witness $\chi_z$
expressed by condition~(\ref{in1}). Therefore the critical temperature
revealed by the witness $\chi_z$ also applies to EW2, and in the following
it is jointly denoted by $ T_c^{exp} $.

Since the magnetic susceptibility of the system $\rm Na_2V_3O_7$ is
available from Sec.~III, we next calculate the critical temperature
$T_c^{exp}$ signaled by $\chi_z$. Application of criterion~(\ref{in1})
implies that entanglement is present in the system, when the inverse
magnetic susceptibility per volume satisfies the inequality
\begin{eqnarray}
\label{in15}
 \chi _z^{ - 1}  > 6\frac{T}{C}\;,
\end{eqnarray}
where the concrete value of $ C = \left( {g\mu _B } \right)^2 N/k $ for the
material $\rm Na_2V_3O_7$ is given by $C\approx 5.58$~ emu~K/mol. The
equality in (\ref{in15}) is represented by the dotted line in Fig.~2. The
intersection point of the solid curve (calculated inverse susceptibility
versus $T$) and the dotted line provides the estimate of the critical
temperature $ T_c^{\exp} \approx 365$~K. One may note that experimental data
was measured up to $315$~K and therefore the estimation was obtained by the
calculated curve, which is not so reliable. However, data of inverse
susceptibility is larger than the right-hand side of~(\ref{in15}) up to
$T=315$~K, thus $T_c^{exp}$ is definitely higher than $315$~K, and therefore
the experimental data in this temperature range cannot be described fully
without taking into account entanglement.

Now relying on the assumption that the partially frustrated
model~(\ref{in14}) describes well the real physical system, we calculate a
lower bound on the presence of entanglement, witnessed by EW1 of
criterion~(\ref{in9}). If for any two sites $(i,j)$ of the ring of nine
qubits the inequality $ \left| {\left\langle {{\bf S}_i {\bf S}_j }
\right\rangle } \right| > 1/4$ holds, then the state of the pair $(i,j)$,
described by the reduced density matrix of those two, is entangled.
Furthermore, regarding the result of Ref.~\cite{WZ02}, if $ \left\langle
{{\bf S}_i {\bf S}_j } \right\rangle$ is negative we can state that the
quantum correlation between the $i$th and $j$th sites is due to pairwise
entanglement.
\begin{figure}
\vspace{-1cm}
\includegraphics[width=8.7cm]{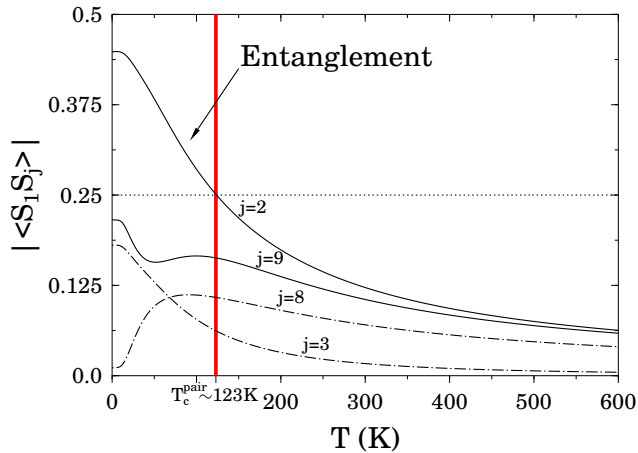}\vspace*{-3cm}
\vspace{1.5cm} \caption{Temperature dependence of the absolute value of the
two-site spin correlation function $\left| {\left\langle {{\bf S}_1 {\bf
S}_j } \right\rangle } \right|$, for neighboring ($j=2,9$) and second
neighboring  ($j=8,3$) sites. The solid curves correspond to nearest
neighboring sites, while the chain curves represent spin-spin correlations
between second neighbors. The horizontal dotted line represents the value of
$1/4$ for the EW1. The intersection point of this line with the solid curve
corresponding to $j=2$ gives the critical temperature $ T_c^{pair} \approx
123$~K. The critical temperature is designated by a vertical solid line
which defines the temperature range (left from the solid line) with pairwise
entanglement in the system $\rm Na_2V_3O_7$.} \label{structure}
\end{figure}
%===========================================================================
Fig.~3 shows the $ \left| {\left\langle {{\bf S}_1 {\bf S}_j } \right\rangle
} \right|$ correlation functions versus $T$ calculated in the temperature
range $T=0-600$~K where the labels $j=2,9$ and $j=8,3$ correspond to the
first and second neighbors of site 1, respectively (see Fig.~1). We do not
show spin-spin correlation functions for the third and fourth neighboring
sites because of their small values. By the calculation of the correlation
functions we used exactly the same model parameters as by the calculation of
the inverse magnetic susceptibility in Sec.~III so that the results in
Fig.~2 and Fig.~3 would be consistent. Due to the symmetry of the model
system, there are two inequivalent neighboring $(j=2,9)$ and second
neighboring $(j=8,3)$ sites. Numerically we obtained that apart from the
case $ \left\langle {{\bf S}_1 {\bf S}_3 } \right\rangle$ (i.e. $j=3$), all
the other correlation functions (i.e. $j=2,9,8$) are negative in the whole
range of $T=0-600$~K. On the other hand, it can be noticed in Fig.~3 that
the only curve which intersects the value of ${1/4}$, represented by the
horizontal dotted line, is the function $\left| {\left\langle {{\bf S}_1
{\bf S}_2 } \right\rangle } \right|$. Since $ \left\langle {{\bf S}_1 {\bf
S}_2 } \right\rangle$ in the whole temperature range is negative, it implies
that the estimated value of the critical temperature $ T_c^{pair} \approx
123$~K refers to bipartite entanglement. According to the above analysis we
conclude that EW1 is able to detect pairwise entanglement between six first
neighboring sites, namely between $(i,j)=(1,2)$, $(2,3)$, $(4,5)$, $(5,6)$,
$(7,8)$, $(8,9)$.

We have also calculated the spin-spin correlation for the $T=0$ ensemble,
where the ground state density matrix was assumed to be an equal mixture of
all eigenstates, corresponding to the fourfold degenerate ground state. For
the next nearest neighboring (NNN) spin pairs the values ${\left\langle
{{\bf S}_1 {\bf S}_8 } \right\rangle=-0.011}$ and ${\left\langle {{\bf S}_1
{\bf S}_3 } \right\rangle=+0.1807}$ were obtained. As it can be observed in
Fig.~3, these are just the limiting values of these NNN spin-spin
correlation functions by $T=0$, which indicates smooth spin-spin correlation
functions near $T=0$. The large difference in the values of ${\left\langle
{{\bf S}_1 {\bf S}_8 } \right\rangle}$ and ${\left\langle {{\bf S}_1 {\bf
S}_3 } \right\rangle}$ implies strong partial frustration in the system due
to the large coupling $J_2$, which diminishes the spin-spin correlation
between sites $1$ and $8$ at $T=0$. Although this effect may reduce the
critical temperature pertaining to the pairwise entanglement, qualitatively
the result obtained in this section for the $\rm Na_2V_3O_7$ is consistent
with the result for the symmetric trimer we arrived at Sec.~II. In the
latter case the bound revealed by EW2 ($\left| {\left\langle {{\bf S}_1 {\bf
S}_2 } \right\rangle } \right|=1/8$) was half of the bound corresponding to
EW1 ($\left| {\left\langle {{\bf S}_1 {\bf S}_2 } \right\rangle }
\right|=1/4$). On the other hand, we can calculate the power of EW2 with
respect to EW1 in the system $\rm Na_2V_3O_7$ as well. By eye inspection of
the spin-spin correlation functions in Fig.~3, the value $\left|
{\left\langle {{\bf S}_1 {\bf S}_2 } \right\rangle } \right|$ by the
critical temperature $T_c^{exp}=365$~K approximately yields $\left|
{\left\langle {{\bf S}_1 {\bf S}_2 } \right\rangle } \right|=1/10$. This
demonstrates the fact that despite of the frustration, there exists
entanglement in the system $\rm Na_2V_3O_7$ beyond the pairwise one. The
residual thermal entanglement, stored in three-spin, four-spin, and
higher-spin correlations, could not be detected by the $\left| {\left\langle
{{\bf S}_1 {\bf S}_2 } \right\rangle } \right|=1/4$ bound of EW1, but was
revealed by the macroscopic witness $\chi_z$ (based on EW2).

\smallskip

\noindent\textbf{V. Conclusion} \\In conclusion we have shown that the
nanotubular system $\rm Na_2V_3O_7$ is an ideal candidate to demonstrate
thermal entanglement at the range of room temperature. The magnetic
susceptibility as a macroscopic entanglement witness, proposed in
Ref.~\cite{WVB05a} (and based on EW2 in criterion~(\ref{in12})) gives the $
T_c^{exp} \approx 365$~K lower bound on the critical temperature beyond
which entanglement disappears, while the calculation of EW1 (of
criterion~(\ref{in9})) for the model Hamiltonian of the $\rm Na_2V_3O_7$
system revealed pairwise entanglement up to the temperature $T_c^{pair}
\approx 123$~K. This result indicates that pairwise entanglement is not the
sole entanglement in the system, and the magnetic susceptibility $\chi_z$ is
able to exhaust a probably large portion of residual part of the total
entanglement. However we have to mention that the total entanglement in this
particular material does not extend to the whole system, but rather is
confined to the rings of the tube. Although macroscopic observables, such as
magnetic susceptibility, can be explained only on the ground of entangled
states, the entangled state itself is not macroscopic but shared among nine
spins within a cluster. We emphasize that while the calculation of the
temperature bound $T_c^{pair}$ for the pairwise entanglement depends on the
validity of the model of the real physical system, the critical temperature
$T_c^{exp}$ revealed by the entanglement witness $\chi_z$ does not require
any assumption about the explicit values of the exchange-coupling parameters
of the system. Moreover, it does not rely on the existence of independent or
weakly coupled spin clusters in the material, thus the result for
$T_c^{exp}$ is reliable even if our model assumptions do not hold exactly.
The existence of large exchange coupling parameters are of great importance
in order to demonstrate quantum entanglement at high temperatures in the
solid. Since this requirement is fulfilled for a large class of materials,
such as some low-dimensional \cite{LGG03} or finite \cite{HBM04} quantum
spin systems, we believe that entanglement as a crucial resource for quantum
information processing in various types of solids might be harnessed at room
temperature range as well.

\noindent\textbf{Acknowledgement} \\
The authors wish to thank Zs.~Gul\'acsi for helpful discussions.


\begin{thebibliography}{99}

\bibitem{Peres93} A. Peres, "Quantum Theory: Concepts and Methods",
(Kluwer Academic Publishers, 1993).

\bibitem{Nielsen00} M. A. Nielsen and I. L. Chuang, "Quantum
Computation and Quantum Information", (Cambridge University Press,
2000).

\bibitem{Ghosh03} S. Ghosh, T. F. Rosenbaum, G. Aeppli, and S. N.
Coppersmith, Nature {\bf 425}, 48 (2003); V. Vedral, Nature {\bf 425}, 28
(2003).

\bibitem{BVZ04} \v C. Brukner, V. Vedral, and A. Zeilinger, Phys. Rev. A {\bf
73}, 012110 (2006).

\bibitem{HHH96} M. Horodecki, P. Horodecki,
and R. Horodecki, Phys. Lett. A {\bf 223}, 1 (1996).

\bibitem{BFS63} L. Berger, S. A. Friedberg, and J. T. Schriemf,
Phys. Rev. {\bf 132}, 1057 (1963).

\bibitem{BT05} I. Bose and A. Tribedi, Phys. Rev. A {\bf 72}, 022314 (2005).

\bibitem{Toth05} G. T\'oth, Phys. Rev. A {\bf 71}, 010301(R) (2005).

\bibitem{Wang05} X. Wang, Phys. Lett. A {\bf 334}, 352 (2005).

\bibitem{DDB04} M. R. Dowling, A. C. Doherty and S. D.
Bartlett, Phys. Rev. A {\bf 70}, 062113 (2004).

\bibitem{BV04} \v C. Brukner and V. Vedral, e-print quant-ph/0406040.

\bibitem{WBSL04} L.-A. Wu, S. Bandyopadhyay, M. S. Sarandy, and D. A. Lidar,
Phys. Rev. A {\bf 72}, 032309 (2005).

\bibitem{WVB05b} M. Wie\'sniak, V. Vedral, and \v C. Brukner,
e-print quant-ph/0508193.

\bibitem{WVB05a} M. Wie\'sniak, V. Vedral, and \v C. Brukner,
New J. Phys. {\bf 7}, 258 (2005).

\bibitem{Millet99} P. Millet, J. Y. Henry, F. Mila, and J.
Galy, J. Solid State Chem. {\bf 147}, 676 (1999).

\bibitem{Dasgupta04} T. Saha-Dasgupta, R. Valent\'\i, F. Capraro, C. Gros,
Phys. Rev. Lett. {\bf 95}, 107201 (2005).

\bibitem{Wang02} X. Wang, Phys. Rev. A {\bf 66}, 044305 (2002).

\bibitem{HV05} B. C. Hiesmayr and V. Vedral, e-print
quant-ph/0501015.

\bibitem{Hofmann03}
H. F. Hofmann and S. Takeuchi, Phys. Rev. A {\bf 68}, 032103
(2003).

\bibitem{WZ02} X. Wang and P. Zanardi, Phys. Lett. A {\bf 301}, 1 (2002).

\bibitem{Wootters98} W. K. Wootters, Phys. Rev. Lett. {\bf 80}, 2245 (1998).

\bibitem{CKW00} V. Coffman, J. Kundu, and W. K. Wootters, Phys. Rev. A {\bf 61},
052306 (2000).

\bibitem{LNMKM04} A. L\"uscher, R. M. Noack, G. Misguich, V. N. Kotov, and F.
Mila, Phys. Rev. B {\bf 70}, 060405(R) (2004).

\bibitem{GRMOMM} J. L. Gavilano, D. Rau, S. Mushkolaj, H. R. Ott, P.
Millet, and F. Mila, Phys. Rev. Lett. {\bf 90}, 167202 (2003).

\bibitem{LGG03} P. Lemmens, G. G\"untherodt, and C. Gros,
Phys. Rep. {\bf 375}, 1 (2003).

\bibitem{HBM04} J. T. Haraldsen, T. Barnes, and J. L. Musfeldt,
Phys. Rev. B {\bf 71}, 064403 (2005).

\end{thebibliography}
\end{document}